\documentclass[preprint]{elsarticle}
\usepackage{natbib}
\usepackage{amsthm}
\usepackage{amsmath}
\usepackage{graphicx}
\usepackage{xcolor}
\usepackage{caption}
\usepackage{subcaption}
\usepackage{tikz}
\usepackage{amssymb}
\usepackage{xurl}
\graphicspath{{/images/}}

\newtheorem{theorem}{Theorem}
\newtheorem{lemma}{Lemma}
\newtheorem{corollary}{Corollary}
\newtheorem{definition}{Definition}

\usepackage{todonotes}

\begin{document}
	
	\begin{frontmatter}	
		\title{Social Welfare and Price of Anarchy in \\ Preemptive Priority Queues}
		\author{Jonathan Chamberlain\corref{cor1}}
		\ead{jdchambo@bu.edu}
		\author{David Starobinski}
		\ead{staro@bu.edu}
		\address{Department of Electrical and Computer Engineering, Boston University, 8 St Mary's Street, Boston MA 02215}
		
		\cortext[cor1]{Corresponding author}
		
		\begin{abstract}
		Consider an unobservable $M|G|1$ queue with preemptive-resume scheduling and two priority classes. Customers are strategic and may join the premium class for a fee. We analyze the resulting equilibrium outcomes, equilibrium stability, and social welfare. We find that for service distributions with coefficient of variation greater than 1, there exists a unique and stable mixed equilibrium at low loads. We also establish a tight bound on the price of anarchy, which is $4/3$. 
		
		\end{abstract}
		
		
	\end{frontmatter}
	
	\section{Introduction}
	\label{section:introduction}
	
    Queuing models based on priority scheduling have been applied to a variety of contexts. These include transmission of multimedia traffic over a network \citep{walraevens2000performance}, management of hospital beds and ambulances \cite{Peköz2002}, and managing the smart grid \citep{6243562}. All of these examples concern themselves with non-preemptive models. Yet, many other important applications, such as high performance computing \cite{Cook} and scheduling of cloud containers \cite{OpenShiftPriority}, make use of \emph{preemption}. As far as we are aware, however, there are relatively few studies of strategic customer behavior in preemptive queues. 
	
    While equilibria and social welfare within preemptive priority queues are considered in works such as \citep[pp 83-85]{ToQueueorNottoQueue} \cite{Shi:2019:AEB:3312662.3312685} \cite{CoverageCoarsnessClassification}, the analyses either implicitly or explicitly assume that the $M|M|1$ regime is in effect. In particular, under a model in which customers may pay a fee to join the higher priority class such as in \citep[pp 83-85]{ToQueueorNottoQueue}, it is asserted that a mixed equilibrium state will never be stable, partly because if one exists it will not be the sole possible equilibrium state. In our paper, we show that this result does not extend to general service distributions.
    
    Specifically, we consider an $M|G|1$ queue with preemptive-resume scheduling discipline (i.e., preempted jobs resume from the point where they are interrupted), and two priority classes. Customers have the option on entering the queue to purchase access to a \emph{premium} class, or otherwise remain in the \emph{ordinary} class. 
    
    We use the standard notation $\lambda$ to denote the mean arrival rate, $\mu$ for the mean service rate, and $\rho = \lambda/\mu$ for the traffic load. In addition, we use $C$ to denote the cost to join the premium class, $\phi$ to denote the fraction of customers joining the premium class, and define a variance parameter $K$ such that the second moment of service equals $K/\mu^2$.

    Under the model of an unobservable queue~\citep[pp 22, 53]{ToQueueorNottoQueue}, we analyze the equilibrium outcomes, stability of the equilibria, and social welfare of the system. We show that the results are influenced both by the traffic load and the second moment of the service distribution. In particular, we show that a stable mixed equilibrium exists at sufficient low traffic load if $K>2$ (i.e., the coefficient of variation is greater than 1). We further show that the price of anarchy of the queue is bounded by $4/3$. 
    These results stand in  sharp contrast to the non-preemptive case where $K$ bears no impact on the equilibrium outcomes and the price of anarchy is always 1.
    
	
	
	\section{Equilibrium Analysis}
	\label{section:eqstates}
	
	We are interested in the existence and stability of equilibria. In this model, the provider fixes the cost $C$ to join the premium class prior to admitting customers. Thus, equilibria states are characterized by the fraction $\phi$ of customers who join the premium class. This results in three possible equilibria types:
	\begin{enumerate}
		\item \emph{Everyone joins} the premium class, i.e. $\phi = 1$. 
		\item \emph{No one joins} the premium class, i.e. $\phi = 0$.
		\item \emph{Some join} the premium class, i.e. $\phi \in (0,1)$.  
	\end{enumerate}
	Equilibria of the first two types are \emph{pure} strategies, as all customers make the same decision. Equilibria of the third type are \emph{mixed} strategies, as a customer who is indifferent will join the premium class with probability $\phi$ and remain in the ordinary class with probability $1-\phi$.
	
    To show existence of possible equilibria, we must define where the customer will be indifferent between their options. As the customer chooses between the premium or ordinary class, they are indifferent if the costs of joining each class are identical. By assumption, customers are statistically identical, thus we need only consider the cost of waiting in the queue. By extension, WLOG we may assume a customer's cost of waiting in the queue equals the time spent waiting as the customers will have identical value on their time spent waiting. Letting $E[W_p]$ and $E[W_o]$ be the expected wait time in the queue as a member of the premium and ordinary classes respectively, a customer is indifferent if the following holds:
    \begin{equation*}
	   E[W_p] + C = E[W_o]. 
	\end{equation*}
    As the wait times depend on the fraction of customers in each class, we can relate a equilibrium strategy $\phi \in [0,1]$ to the cost which leads to that equilibrium by applying the formula for expected wait time in an $M|G|1$ priority-resume (PR) queue: \cite[p.175]{TheoryofScheduling}:
	\begin{equation}
	C(\phi) \triangleq E[W_o] - E[W_p] = \frac{K\rho + (2-K)\phi\rho(1-\rho)}{2\mu(1-\rho)(1-\phi\rho)}.
	\label{eq:C(phi)}
	\end{equation}
	
	 We first evaluate the behavior of $C(\phi)$ as $\phi$ increases from 0 to 1, to show how the relative costs of waiting changes as more customers attempt to join the premium class. 
	
	\begin{lemma}
		The function $C(\phi)$, defined in Equation \eqref{eq:C(phi)}, behaves as follows with respect to $\phi$:
		\begin{enumerate}
			\item If $K > 2$ and $\rho < (K-2)/(2K-2)$, $C(\phi)$ is monotone decreasing.
			\item Else, if $K > 2$ and $\rho = (K-2)/(2K-2)$, $C(\phi)$ is constant valued.
			\item Otherwise, $C(\phi)$ is monotone increasing.
		\end{enumerate}
		\label{lemma:System States}
	\end{lemma}
	
	The proof is obtained by computing the derivative $C'(\phi)$ and determining the conditions for which it is positive, negative, or zero. It turns out that the sign of the derivative is determined by the sign of its numerator, which is constant with respect to $\phi$. Therefore, $C(\phi)$ must be monotone or constant, the exact behavior depending on the values of $K$ and $\rho$, and the rest follows. 
	
	As $C(\phi)$ must be monotone or constant, $\min{C(\phi)}$ and $\max{C\phi)}$ will be well defined quantities. Further, each of these will be equal to either $C(0)$ or $C(1)$ depending on the exact behavior of $C(\phi)$. If $C(\phi)$ is monotone, there is a unique solution to $C(\phi) = C$,  which is 
	\begin{equation}
	        \phi_e = \frac{2\mu C(1-\rho) - K\rho}{\rho(1-\rho)(2\mu C+2-K)}.
	        \label{eq:phi_e}
	    \end{equation}

	Thus, given a cost $C$, we determine the existence of possible equilibria by relating $C$ to $C(\phi)$. When evaluating the stability of any such possible equilibria, we apply the Evolutionary Stable Strategy definition from \cite{EvolutionTheoryofGames}:
	
	\begin{definition}
		A strategy adopted by a population which cannot be invaded by an initially rare strategy is said to be an \emph{Evolutionary Stable Strategy (ESS)}. That is, if strategy $\phi^*$ is ESS, then no other equilibrium strategy $\phi^{**}$ exists such that $\phi^{**}$ is a best response to $\phi^*$. 
	\end{definition}

We next present the main results of this section:	
	
	\begin{theorem}
	   
	    \noindent The equilibria of a two-class $M|G|1$-PR queue have the following structure:
	    \begin{enumerate}
	        \item If $C < \min C(\phi)$, \emph{everyone joins} is the unique equilibrium. 
	        \item If $C > \max C(\phi)$, \emph{no one joins} is the unique equilibrium.
	        \item If $\min C(\phi) < C < \max C(\phi)$ and $C(\phi)$ is monotone decreasing, a \emph{some join} equilibirum with $\phi_e$ customers in the premium class is the unique equilibrium. 
	        \item If $\min C(\phi) < C < \max C(\phi)$ and $C(\phi)$ is monotone increasing, the \emph{everyone joins} equilibrium, \emph{no one joins} equilibirum, and \emph{some join} equilibrium with $\phi_e$ customers are all possible equilibria.
	    \end{enumerate}
	    Pure equilbibria are always ESS. The mixed equilibrium is ESS if and only if it is unique.
	    \label{thm:eqstates}
	\end{theorem}
	
\begin{corollary}
By Lemma 1 and Theorem 1, a unique mixed equilibrium exists if and only if  $K > 2$ and $\rho < (K-2)/(2K-2)$. Furthermore, this equlibrium is ESS.        
\end{corollary}

	We next sketch the proof of Theorem \ref{thm:eqstates}. The first two cases follow trivially from the fact that if $C$ is smaller than the minimum (or respectively, greater than the maximum) joining the premium class will cost less (resp., more) than remaining in the ordinary class regardless of how many customers are in the premium class. Therefore, the only possible equilibria is the \emph{everyone joins} (resp., \emph{no one joins}) equilibrium.
	
	In the third case, $\phi_e$ being a possible equilibrium follows by definition. That no others are possible follows from $C(\phi)$ being monotone decreasing: if more customers join the premium class than under equilibrium, then it is cheaper to remain in the ordinary class, and vice versa. Hence, a deviation from $\phi_e$ has a best response of pushing the system back to $\phi_e$, and no other equilibrium is possible.
	
	In the fourth case, $\phi_e$ is again a possible equilibrium state by definition. However, because $C(\phi)$ is monotone increasing, if more customers join the premium class than under equilibrium, then it is cheaper to join the premium class. The reverse is also true if more customers join the ordinary class than under equilibrium. This results in the system reaching a pure state if deviating from $\phi_e$. These pure states are also possible equilibria as once in a pure state, it costs more to attempt to join the opposite class of what all other customers have chosen.
	
	The ESS criteria follows as a corollary from the previous assertions. In the first three cases, deviating from the equilibrium is never a best response. In the fourth case,  pure equilibria are possible by deviating from the mixed equilibrium. 
	
	\section{Social Welfare Analysis}
	\label{section:socialoptimal}
	
    We now shift our analysis to the social welfare and attendant price of anarchy~\citep{6545289}. Social welfare is defined in terms of the utilities of the customers and the provider. However, here customers are statistically identical, and the preemption policy is work-conserving. Further, the cost $C$ to join the premium class is a transfer from customers to the provider. As a result, the social welfare only varies based on the costs of waiting in the queue for service. Thus we maximize the social benefit by minimizing overall wait times. We derive the expected average wait time $E[W]$ from the expected wait times in each priority class as follows:
	
	\begin{equation}
	E[W] = \frac{\rho\left(K-2\phi\rho+(2-K)\phi(1-\phi(1-\rho))\right)}{2\mu(1-\rho)(1-\phi\rho)}.
	\label{eq:Average Wait Time PR}
	\end{equation}

	\begin{lemma}
		Let $\phi^*$ be defined as follows:
		\begin{equation}
	        \phi^* \triangleq \frac{1 - \sqrt{1-\rho}}{\rho}.
	        \label{eq:phi star}
	    \end{equation}
		In this model, the socially optimal states depend on the value of $K$ as follows:
		\begin{enumerate}
			\item If $K < 2$, the socially optimal states are $\phi = 0$ and $\phi = 1$;
			\item If $K = 2$, all states $\phi \in [0,1]$ are socially optimal;
			\item If $K > 2$, the socially optimal state is $\phi = \phi^*$.
		\end{enumerate}
		\label{lemma:PR Social Optimal State}
	\end{lemma}
	
	To prove this, we compute the derivative of $E[W]$ with respect to $\phi$. In doing so, we find that the sign of the derivative flips when $\phi = \phi^*$. If $K < 2$, the sign of the derivative flips from positive to negative at $\phi^*$, thus $\phi^*$ results in the maximum possible wait time, and conversely $\phi = 0$ and $\phi = 1$ result in the minimum wait time. If $K = 2$, then the derivative is 0 for all $\phi$, thus the wait time is constant with respect to $\phi$ and all states are optimal by default. If $K > 2$, the sign of the derivative flips from negative to positive at $\phi^*$ and therefore $\phi^*$ results in the minimum possible wait time.
	
	As $K$ is defined in terms of the second moment, this means that if variance in service is less than that of exponential, the social welfare is maximized when all customers join the same class. If variance is greater than that of exponential, the social welfare is maximized when a specific fraction $\phi^*$ of customers join the premium class. If the service distribution is exponential however, then it does not matter how many customers join the premium class, as the choice to join or not does not impact the overall welfare. This can ultimately be shown to result from the relation between the expected time of service of a preempting customer ($1/\mu$) and the expected residual time of service of the preempted customer $\left(K/(2\mu) \right)$. 

	\subsection{Price of Anarchy}
	\label{subsection:poa}

     The Price of Anarchy ($PoA$) is a measure of the loss of optimality resulting from a lack of cooperation. This is defined by comparing the socially optimal state to the equilibrium state which leads to the highest social cost \citep{6545289}. As the social welfare depends solely on $E[W]$, we define the $PoA$ for our system in terms of the costs of waiting:
     
	\begin{definition}
		Let $E \subset [0,1]$ be the set of possible equilibria for fixed cost $C$ and traffic load $\rho$. The Price of Anarchy ($PoA$) is defined as the following ratio:
		\begin{equation*}
		PoA = \frac{\underset{\phi \in E}{\max}E[W]}{\underset{\phi \in [0,1]}{\min}E[W] }.
		\end{equation*}
		\label{def:PoA}
	\end{definition} 
	
	As possible equilibria and the socially optimal state depend on  $K$, $\rho$, and $C$, so must the $PoA$. As a result, we aim to show that an upper bound exists on the $PoA$, such that there is a clearly defined \emph{worst case} scenario which cannot be exceeded. 
	
	\begin{theorem}
	    The price of anarchy of a two-class $M|G|1$-PR queue is bounded from above by $4/3$.
	    \label{thm: PoA upper bound}
	\end{theorem}
	
	To prove that such an upper bound exists, we assume that given arbitrary but fixed $\rho$ and $K$, $C$ is set such that the state $\phi$ which leads to the largest possible average wait time $E[W]$ is in the set of possible equilibria. Per Lemma \ref{lemma:PR Social Optimal State}, we note that if $K<2$, the socially optimal states are $\phi \in \{0,1\}$. Using the proof of the lemma, it is straightforward to show that $\phi^*$ is the corresponding worst-case state. This results in  
	\begin{equation}
	    PoA = \frac{(2-K)\left(2-2(1-\rho)^\frac{3}{2}-3\rho\right)}{K\rho^2} + \frac{2}{K}.
	    \label{eq: PoA K<2}
    \end{equation}
    If $K = 2$, then all states are socially optimal, and thus so are all possible equilibria states. It then follows that if $K = 2$, the $PoA$ is equal to 1, regardless of the value of $\rho$. 
    
    If $K > 2$, then the socially optimal state is $\phi = \phi^*$, and the worst case state is $\phi \in \{0,1\}$. Thus, we have a $PoA$ which is the reciprocal of expression in Equation \eqref{eq: PoA K<2}:
    \begin{equation}
	        PoA = \frac{K\rho^2}{(2-K)(2-2(1-\rho)^\frac{3}{2}-3\rho)+2\rho^2}.
	        \label{eq: PoA K>2}
	\end{equation}
	
	Thus, to determine an the upper bound on the price of anarchy, we determine the suprema of Equations \eqref{eq: PoA K<2} and \eqref{eq: PoA K>2} given $\rho \in (0,1)$, and the respective bounds on $K$ for each equation. 
	We observe that Equation \eqref{eq: PoA K<2} will be maximized when $K=1$; evaluating with respect to $\rho$ we find that the $PoA$ when $K < 2$ is bounded from above by $5/4$, as $\rho$ approaches 0. Evaluating Equation \eqref{eq: PoA K>2}, we find that it too will be maximized as $\rho$ approaches 0, resulting in a bound of
	\begin{equation*}
	    PoA =\frac{4K}{2+3K}.
	\end{equation*}
	And as $K$ approaches infinity, this quantity is bounded from above by $4/3$. Thus, regardless of the values of $K$, $C$, or $\rho$, the price of anarchy is never greater than $4/3$. This bound is reached as $\rho$ approaches 0. Thus, low traffic loads lead to the greatest cost from a lack of cooperation. Conversely, as $\rho \to 1$, the $PoA$ approaches 1.

    
    \section{Comparison to the Non-Preemptive (NP) Queue}
    
    We now briefly contrast the results to the situation where customers are offered the ability to purchase access to the premium class, but no customers may be preempted while in service. Based upon the formula for waiting in an $M|G|1$ queue with no preemption~\cite[p.164]{TheoryofScheduling}, one can derive the resulting cost function $C(\phi)$ and expected average wait time $E[W]$ as follows:
    \begin{equation*}
        \begin{split}
            C(\phi) & = \frac{K\rho^2}{2\mu(1-\rho)(1-\phi\rho)}, \\
            E[W] & = \frac{K\rho}{2\mu(1-\rho)}.
        \end{split}
    \end{equation*}
    
    We immediately note that $E[W]$ is constant with respect to $\phi$, thus the average wait time does not depend on the number of customers in each class. This results in all states being socially optimal and the $PoA=1$ by default, regardless of which equilibria are possible. Evaluating $C(\phi)$, we find that in a non-preemptive queue, the following equilibria structure prevails:
    \begin{itemize}
		\item If $C < C(0)$,  \emph{everyone joins} is the only possible equilibrium.
		\item If $C > C(1)$,  \emph{no one joins} is the only possible equilibrium.
		\item If $C(0) < C < C(1)$, there are three possible equilibria: \emph{everyone joins}, \emph{no one joins}, and a \emph{some join} with $\phi = 1/\rho - (K\rho)/(2\mu C(1-\rho))$. 
	\end{itemize}
	
	The pure equilibria are ESS, while the \emph{some join} will never be ESS. Thus, in contrast to the preemptive queue, customers will tend towards all joining the same class no matter what the service variance or traffic load are. Indeed, the service variance will only impact the cost that can be charged to join the premium class, but it has no impact on the social welfare. 
    
	\section{Conclusions}
	In this paper we analyzed the equilibrium outcomes of a two-class $M|G|1$-PR queue where customers can purchase access to the higher priority class for a fee. We find that if the variance in service is greater than that of the exponential distribution and the traffic load is sufficiently low, then there exist conditions under which a stable mixed equilibrium is possible. This is not possible otherwise (e.g., for exponential and deterministic service times). 
	
We also conducted a social welfare analysis and showed that a mixed state is socially optimal if service variance is sufficiently high (i.e., $K>2$). Thus, it is not the case generally that the system is always best off when all customers join one class or the other, as might be observed when looking at more deterministic systems. In any event, the resulting price of anarchy is bounded by $4/3$. As the traffic load increases to 1, the price of anarchy tends to 1. Finally, we observe that the social welfare in the $M/G/1$-PR queue can either be smaller or larger that of the $M/G/1$-NP queue. However, the former case is likelier because only in the $M/G/1$-PR queue a mixed equilibrium is stable, and this can only be reached when $K>2$ (i.e., when a mixed equilibrium leads to a higher social welfare than the pure equilibria).


\section*{Acknowledgements}
This work was  partially supported by NSF grants 1717858 and 1908087.
	
	\section*{Bibliography}
	\bibliographystyle{elsarticle-num}
	\bibliography{bibfile}
	
\end{document}